\documentclass[11pt,twoside]{article}
\usepackage{asp2010}

\resetcounters

\begin{document}

\title{Parker Winds Revisited: An Extension to Disk Winds}
\author{Timothy R. Waters,$^1$ Daniel Proga,$^{1,2}$}
\affil{$^1$Department of Physics, University of Nevada, Las Vegas,
NV 89154}
\affil{$^2$Princeton University Observatory, Peyton Hall, Princeton, NJ 08544}

\begin{abstract}
A simple, one-dimensional dynamical model of thermally driven disk winds, one in the spirit of the original Parker (1958) model, is presented.  We consider two different axi-symmetric streamline geometries:  geometry (i) is commonly used in kinematic models to compute synthetic spectra, while geometry (ii), which exhibits self-similarity and more closely resembles the geometry found by many numerical simulations of disk winds, is likely unused for this purpose --- although it easily can be with existing kinematic models.  We make the case that it should be, i.e. that geometry (ii) leads to transonic wind solutions with substantially different properties.
\end{abstract}

\section{Introduction}
Developing baseline disk wind models analogous to the spherically symmetric Parker model (Parker, 1958) has proven to be a difficult task.  A major roadblock has been the uncertainty in the streamline geometry.  Another obvious and related difficulty is posed by the fact that accretion disks span many more orders of magnitude in physical size than do stars, and they can host radically different, spatially and temporally variable, thermodynamic environments.  It should come as no surprise then, that despite clear observational evidence of outflows from many systems, identifying the actual driving mechanisms, as well as determining the wind geometry, remains a challenge. 

Studies of disk winds therefore rely heavily on kinematic models in order to quickly explore the parameter space without assuming a particular driving mechanism.  A popular choice of geometry, one that has been used in conjunction with sophisticated radiative transfer simulations to model accretion disk spectra from many systems, including AGN (Sim et al. 2008), is the Converging model --- geometry (i) in Figure~1.  Recent multi-dimensional, time-dependent simulations of a thermally driven wind carried out by Luketic et al. (2010) suggest that the Converging model may not be well-suited for sampling the entire wind, but rather only the inner portions of it.  The outer portion is better approximated by a model in which streamlines emerge at a constant inclination angle $i$ to the midplane (hence the name, the CIA model --- geometry (ii).) 

We have generalized the isothermal and polytropic Parker wind solutions so that they apply to geometries (i) and (ii).  Our solutions amount to a simple \textit{dynamical} disk wind model (see Waters \& Proga 2012).  Rather than positing a velocity law as is done for kinematic models, the purpose of a dynamical model is to impose the physical conditions and solve for the wind velocity as a function of distance along a streamline.  Here we summarize our findings for how the streamline geometry alone can result in winds with substantially different flow properties, limiting our attention to the isothermal case.

\section{Results \& Conclusions}
The long-dashed and solid curves in the plot in Figure~1 depict the steady-state flow properties of a Parker-like disk wind traversing geometries (i) and (ii), respectively.  Specifically, we plot the equivalent nozzle function (denoted $\mathcal{N}$) along a streamline, in units of the gravitational radius.  Also shown are $\mathcal{N}$ for the spherically symmetric (bottom dotted curve) and Keplerian (a radial Parker wind with a Keplerian azimuthal velocity component; topmost dashed-dotted curve) Parker winds.  Revolving $\mathcal{N}$ about the horizontal axis sweeps out the shape of a de Laval Nozzle that yields steady-state flow properties identical to that of the wind; this shape is exponentially dependent on the effective potential and the squared ratio of the local escape velocity to the sound speed (the HEP).  Comparing the throat locations and corresponding magnitudes of $\mathcal{N}$ for geometries (i) and (ii), it is clear that the CIA model has a sonic point distance about twice that of the Converging model (implying a smaller acceleration) and an initial Mach number $\mathcal{M}_o = \mathcal{V}_o/c_s$ that is smaller by nearly an order of magnitude.  Since $\mathcal{M}_o$ is a direct gauge of the mass flux density, the total mass loss rate for a CIA wind will be smaller in general.  These differences all result from the confined expansion area of the CIA model, due to its lack of adjacent streamline divergence.  

Both winds experience a reduced centrifugal force at $i = 60^\circ$ compared to a Keplerian Parker wind, explaining why the latter has a significantly higher initial Mach number.  We can therefore arrive at the result that the mass flux densities of our disk wind models are always bounded from below by that of the spherically symmetric Parker wind and above by that of the Keplerian Parker wind. 

In summary, the different properties of the CIA and Converging models are solely due to geometric effects.  If, for a given HEP and $i$, the resulting velocity profiles were approximated by a beta-law, the parameters $\mathcal{V}_o$ and $\beta$ (the slope) might differ by an order of magnitude.  Kinematic models that make use of a beta-law are therefore sensitive to the type of wind geometry.  The implication is that employing the Converging model may lead to significant overestimates of the flow acceleration if the true streamline geometry more closely resembles the CIA model.  The synthetic line profiles will be affected, especially if the ionization balance of the wind is assumed to depend upon the density or temperature profiles, which significantly differ for these geometries.

\begin{figure}[!ht]
\plottwo{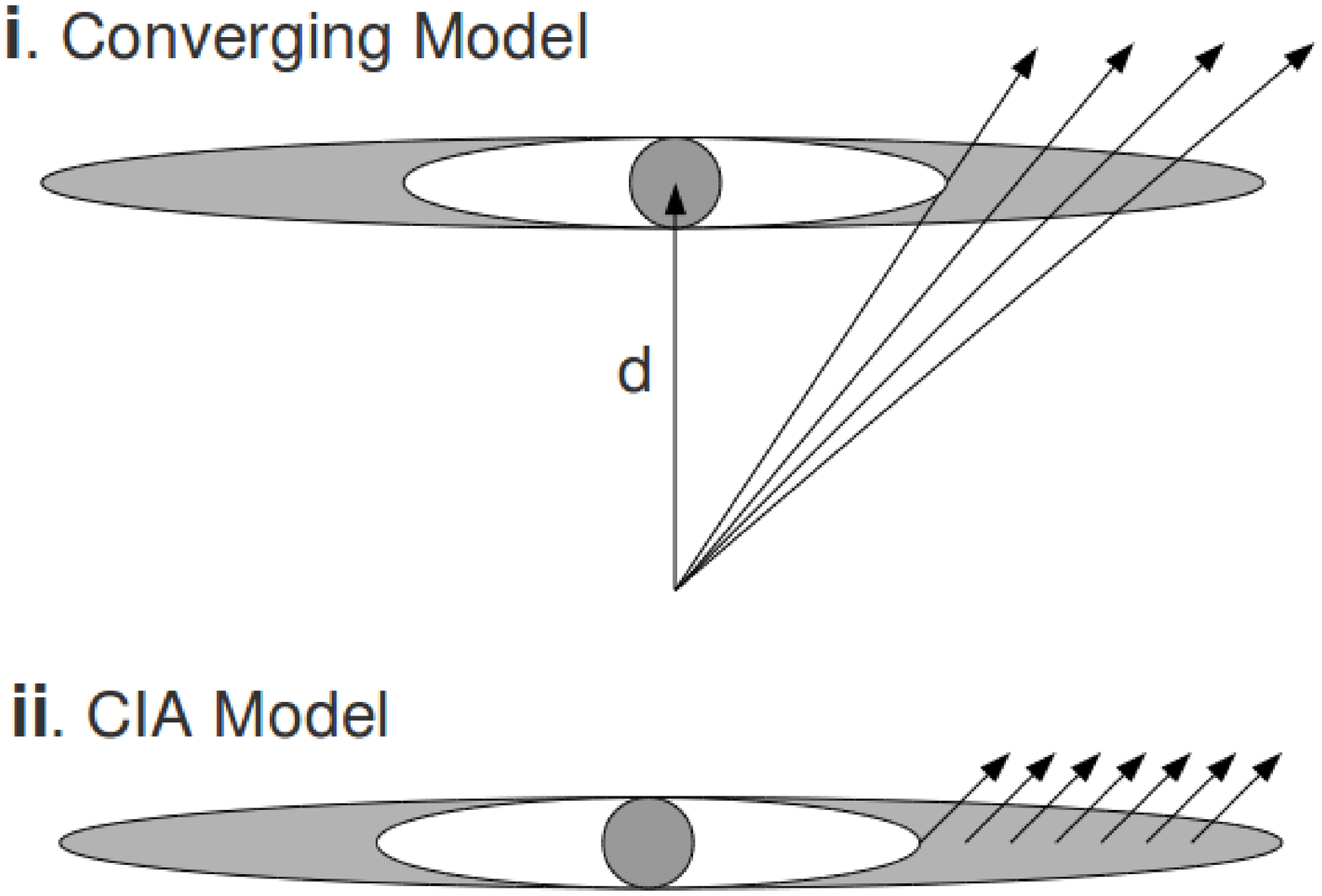}{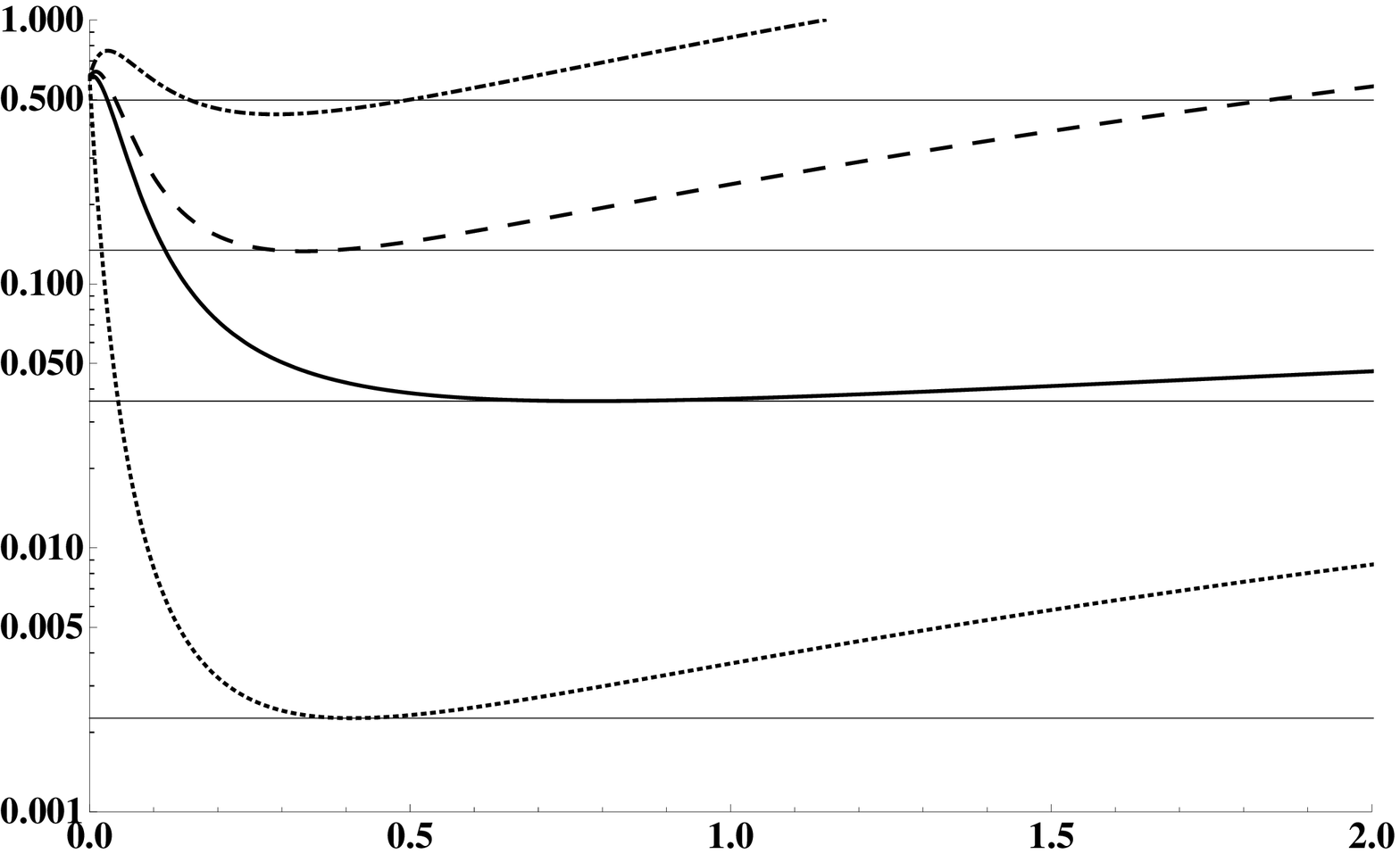}
\caption{ Adjacent streamlines diverge from each other in the Converging model but not in the CIA model.  The plot of equivalent nozzle functions was calculated by taking HEP $=11$ and $i=60^\circ$.  We have normalized $\mathcal{N}$ such that $\mathcal{N} \approx \mathcal{M}_o$ at the nozzle throat; the horizontal lines mark the exact values of $\mathcal{M}_o$.}  
\end{figure}

%\nocite{*}
%\bibliographystyle{asp2010}
%\bibliography{references}

\end{document}